\documentclass[aps,prl,twocolumn,superscriptaddress]{revtex4}
\usepackage{graphicx}
\usepackage{dcolumn}
\usepackage{bm}

\begin{document}

\preprint{APS/123-QED}
\title{Gapped itinerant spin excitations account for missing entropy in the
hidden order state of URu$_2$Si$_2$}
\author{C.~R.~Wiebe}
\email{cwiebe@magnet.fsu.edu} \affiliation{Department of Physics,
Florida State University, Tallahassee, FL 32306-3016, USA}
\affiliation{National High Magnetic Field Laboratory, Florida State
University, Tallahassee, FL 32306-4005, USA}
\author{J.~A.~Janik}
\affiliation{Department of Physics, Florida State University,
Tallahassee, FL 32306-3016, USA} \affiliation{National High Magnetic
Field Laboratory, Florida State University, Tallahassee, FL
32306-4005, USA}
\author{G.~J.~MacDougall}
\affiliation{Department of Physics and Astronomy, McMaster
University, Hamilton, Ontario L8S 4M1, Canada}
\author{G.~M.~Luke}
\affiliation{Department of Physics and Astronomy, McMaster
University, Hamilton, Ontario L8S 4M1, Canada}
\author{J.~D.~Garrett}
\affiliation{Department of Physics and Astronomy, McMaster
University, Hamilton, Ontario L8S 4M1, Canada}
\author{H.~D.~Zhou}
\affiliation{National High Magnetic Field Laboratory, Florida State
University, Tallahassee, FL 32306-4005, USA}
\author{Y.-J.~Jo}
\affiliation{National High Magnetic Field Laboratory, Florida State
University, Tallahassee, FL 32306-4005, USA}
\author{L.~Balicas}
\affiliation{National High Magnetic Field Laboratory, Florida State
University, Tallahassee, FL 32306-4005, USA}
\author{Y.~Qiu}
\affiliation{NIST Center for Neutron Research, Gaithersburg,
Maryland, 20899-5682, USA} \affiliation{Department of Materials
Science and Engineering, University of Maryland, College Park,
Maryland, 20742, USA}
\author{J.~R.~D.~Copley}
\affiliation{NIST Center for Neutron Research, Gaithersburg,
Maryland, 20899-5682, USA}
\author{Z.~Yamani}
\affiliation{CNBC, Chalk River Labs, Chalk River, ON K0J 1J0,
Canada}
\author{W.~J.~L.~Buyers}
\affiliation{CNBC, Chalk River Labs, Chalk River, ON K0J 1J0,
Canada}
\date{\today }

\begin{abstract}
One of the primary goals of modern condensed matter physics is to
elucidate the nature of the ground state in various electronic
systems.  Many correlated electron materials, such as high
temperature superconductors,\cite{Dagotto} geometrically frustrated
oxides,\cite{Greedan} and low-dimensional
magnets\cite{Affleck}\cite{Dagotto2} are still the objects of
fruitful study because of the unique properties which arise due to
poorly understood many-body effects. Heavy fermion metals\cite{Fisk}
- materials which have high effective electron masses due to these
effects - represent a class of materials with exotic properties,
such as unusual magnetism, unconventional superconductivity, and
``hidden order'' parameters. \cite{Chandra} The heavy fermion
superconductor URu$_2$Si$_2$ has held the attention of physicists
for the last two decades due to the presence of a ``hidden order''
phase below 17.5
K.  Neutron scattering measurements indicate that the ordered moment is 0.03 {$\mu_{B}$}%
, much too small to account for the large heat capacity anomaly at
17.5 K.  We present recent neutron scattering experiments which
unveil a new piece of this puzzle - the spin excitation spectrum
above 17.5 K exhibits well-correlated, \emph{itinerant}-like spin
excitations up to at least 10 meV emanating from incommensurate
wavevectors.  The gapping of these excitations corresponds to a
large entropy release and explains the reduction in the electronic
specific heat through the transition.
\end{abstract}

\maketitle

The central issue in URu$_2$Si$_2$ concerns the identification of
the order parameter which explains the reduction in $\gamma$, and
thus the change in entropy, through the transition at 17.5
K.\cite{Chandra} Numerous speculations about the ground state have
been advanced, from quadrupolar ordering,\cite{Silhanek} to
spin-density wave formation,\cite{Mineev} to ``orbital currents" to
account for the missing entropy.\cite{Chandra2} In this letter we
present cold neutron time-of-flight spectroscopy results which shed
some light on the ``hidden order'' in URu$_{2}$Si$_{2}$.  We have
performed experiments above and below the ordering temperature to
measure how the spin excitations evolve. It is clear from our data
that above {T$_{0}$~} the spectrum is dominated by fast,
itinerant-like spin excitations emanating from incommensurate
wavevectors at positions located 0.4a* about the antiferromagnetic
(AF) points. From the group velocity and temperature dependence of
these modes, we surmise that these are heavy quasiparticle
excitations which form below the ``coherence temperature" and play a
crucial role in the formation of the heavy fermion and ``hidden
order'' state.  The gapping of these excitations, which corresponds
to a loss of accessible states, accounts for the reduction in
$\gamma$ through the transition at 17.5 K.

Figure 1 shows the excitation spectrum of URu$_2$Si$_2$ at 1.5 K in
the H00 plane. The characteristic gaps at $\sim$ 2 meV at the AF
zone center (1,0,0) and $\sim$ 4 meV at the incommensurate
wavevectors (0.6,0,0) and (1.4,0,0) have been known for some
time.\cite{Broholm1} The incommensurate wavevector corresponds to a
displacement of $\sim$ 0.4 a* from the AF zone centers (ie. where
h+k+l = an odd integer, and thus forbidden in the BCC chemical
structure). A scenario for this mode-softening at the incommensurate
position was previously described with a model based upon
oscillatory exchange constants between near neighbors (not uncommon
for RKKY type interactions).\cite{Broholm1},\cite{Broholm2}

Figure 2 exhibits our new neutron results at 20 K in the same H00
plane. Above the phase transition, the sharp spin waves evolve into
weak quasielastic spin fluctuations at the zone center (1,0,0) and
strong excitations at the incommensurate positions (1$\pm$0.4,0,0).
We have considered the possibility that these incommensurate
excitations may be due to magnetovibrational scattering.  This can
arise from a shifting of a the phonon excitations at (2,0,0) to
(1$\pm$0.4,0,0) as allowed through the neutron scattering
cross-section for magnetoelastic coupling.\cite{Squires}  However,
with the small moment size of this system, it is improbable that
such a scattering process is being observed.  It was also originally
reported\cite{Broholm1} that the incommensurate excitations were
just quasielastic fluctuations, as constant Q scans on a triple axis
spectrometer resolved a peak at a finite energy of 0.6 THz = 2.5 meV
and a decrease in intensity as a function of energy typical of an
overdamped response.  What was previously unknown was that the
quasielastic fluctuations were only the lower limit of a band of
high-velocity spin excitations that extend well above the upper
limit of the sharp collective spin excitations of the ordered phase.
As the temperature is increased, the (1,0,0) fluctuations decrease
and the incommensurate fluctuations remain approximately constant to
at least 25 K.\cite{Buyers1}

The implications of this discovery are: (1) The incommensurate
excitations have a well-defined structure as a function of Q, and
thus the electrons are highly correlated \emph{above} 17.5 K. This
is completely unexpected for a system of localized moments in a
paramagnetic state, but similar dynamics above {T$_N$}~ (with the
formation of paramagnons\cite{Cr} for example) has been observed
with neutron scattering experiments in other itinerant electron
systems such as chromium and V$_2$O$_3$.\cite{Cr}, \cite{v2o3}  (2)
The dispersion is such that the maximum group velocity is at least a
factor of 2 larger than the maximum group velocity of the
excitations in the ordered state.  A significant restructuring of
the Fermi surface must be responsible for the hidden order state.
(3) The gapping of these strong spin fluctuations below 17.5 K must
provide a considerable portion of the entropy removal at the
transition. These excitations have a structure in reciprocal space
such that they occur at several symmetry-related wavevectors within
the first Brillouin zone.  The amount of phase space occupied by
these excitations is greater than those at the (1,0,0) AF zone
center. Figure 3 shows constant energy cuts of the excitations in
the H0L plane to emphasize this. Note the weak AF fluctuations at
(1,0,0) and (2,0,1) in comparison to the excitations at the
incommensurate positions. Also note how a cone of scattering
develops about the incommensurate positions, while the AF zone
center simply decreases in intensity as energy transfer is
increased. For comparison, a cut is also shown of the spin-wave
spectrum at 1.5 K in the same energy window.  The results at 20 K
suggest a continuum of excitations within a cone of scattering (as
expected for low-lying excitations at the Fermi surface), as opposed
to the sharp excitations which are gapped below 17.5 K.

An estimate of the contribution to the electronic specific heat term
from the removal of these low-energy spin fluctuations can be made
through an analysis of the spectrum at 20 K.  The specific heat of a
cone of fast excitations centered on incommensurate wave vectors and
extending out by an inverse correlation length $\kappa$ =
$\xi^{-1}$, is given by

\begin{equation}
C(T)=\frac{\partial}{\partial T} \frac{v_a}{8
\pi^3}\int_0^{\xi^{-1}} dq 4\pi q^2 \int_0^{E_{max}} dE \rho_0 f(E)
E
\end{equation}

where v$_a$ is the cell volume, $\rho_0$ is the density of states
and f(E) = coth(E/2 k$_{B}$ T).  The upper limit in the energy
integral has been taken to be E$_{max}$ = k$_{B}$T (above this
energy, the specific heat contribution is small since df/dT
approaches zero exponentially).

If we assume the density of states is the inverse of a large damping
in energy $\Gamma$ = c $\xi^{-1}$, then the specific heat becomes

\begin{equation}
C_v = \frac{v_a \xi^{-2}}{3 \pi^2 c} \times {k_B}^2 T
\end{equation}

where c is the characteristic spin wave velocity.  Previous work has
shown that the excitations lie within the HK0 plane around (0.6, 0,
0).\cite{Wiebe}  Note that due to the dispersion of the spin
excitations, the contribution to the heat capacity is
\textit{linear} in temperature, which is exactly the same power law
dependence for the electronic specific heat.  This term arises due
to these fast itinerant spin excitations in URu$_2$Si$_2$ and is the
reason for the enhanced linear component to the specific heat above
T$_0$ = 17.5 K.  This calculation of a gap opening in the spin
excitation spectrum is equivalent to a calculation of the gap
opening in the Fermi surface.  This is demonstrated by the
similarity of the matrix-element-weighted density of spin-wave
states and infrared spectroscopy measurements, which measure charge
excitations.\cite{Bonn}, \cite{Broholm2}  The spin and charge
degrees of freedom are strongly coupled.

A characteristic correlation length and spin wave velocity need to
be extracted from our data to complete this calculation.  Our data
has been fit to a standard formula for paramagnetic scattering from
a system of correlated spins, also known as the Chou
model.\cite{Chou} This has been used to extract correlation lengths
and spin wave velocities for excitations in the superconducting
cuprates, for example.\cite{Mason}, \cite{Rossat}, \cite{Stock} With
this model, S(q, $\omega$) is represented as

\begin{equation}
S(Q,\omega)=\frac{\hbar\omega}{1-e^{\frac{-\hbar\omega}{kT}}}\frac{A}{\kappa^2+q^2}({\frac{\Gamma}{(\hbar\omega\pm\hbar\omega_q)^2+\Gamma^2}})
\end{equation}

where \overrightarrow{q} = \overrightarrow{Q} -
$\overrightarrow{\delta}$ ($\delta$ is the incommensurate
wavevector), $\kappa$ is the inverse of the correlation length
$\xi$, and the frequencies of the damped spin waves are $\omega_q$ =
cq, where c is the spin -wave velocity ($\Gamma$ = $\kappa$c).
Fitting with these parameters yields a correlation length of $\xi$ =
14(2) Angstroms, and c = 45(10) meV Angstroms (see figure 2).  This
value is similar to the Fermi velocity, reported as $\sim$ 8.84 x
10$^3$ m/s (35 meV Angstroms)\cite{Behnia}, illustrating the
itinerant nature of the excitations.  It is interesting to note that
there is a curious intersection of cuprate and heavy fermion physics
with this analysis - the fast excitations seen in the superconductor
La$_{1.86}$Sr$_{0.14}$CuO$_4$ (LSCO), which have an effective mass
near an electron mass, appear similar to the excitations in
URu$_2$Si$_2$, albeit with a higher velocity because of their
lighter effective mass.\cite{Mason}

The contribution of the spin excitations to the specific heat of
with these parameters for URu$_2$Si$_2$ is 220 +/- 70 mJ/(mol K$^2$)
at 20 K. A comparison of this to the data, in Figure 5, shows a good
agreement with a $\gamma$ of 155(5) mJ/(mol K$^2$).  The gapping of
these excitations leads to a dramatic reduction in $\gamma$ to 38(2)
mJ/(mol K$^2$).\cite{Maple}, \cite{Palstra2}  This is naturally
explained through the gap in the spin excitation spectrum, which has
minima at 2 meV and 4 meV.  The average thermal energy below 17.5 K
is less than 2 meV, and thus most of the spin excitations are
inaccessible.  This is the reason for the reduction of $\gamma$
through the phase transition.  The origin of the size of the
specific heat jump at T$_0$ is still unresolved at this point
although its low temperature limit arises from the removal of the
spin contribution to leave a weak remanent ungapped electronic band.
A gap opens in the spin excitation spectrum with a value of $\sim$
110(10) K, but the identity of this ordered state is still a
mystery.\cite{Wiebe} Through these measurements, we do know that the
nature of the transition at {T$_0$~} seems to be dominated by
itinerant electron physics rather than localized crystal field
effects.  This is the reason why URu$_2$Si$_2$ shows a transition
between itinerant and localized behavior at T$_0$, as shown through
thermal conductivity measurements.\cite{Behnia} Indeed, we have seen
no evidence of crystal field excitations up to at least 10 meV
$\sim$ 110 K.  The complicated crystal field schemes which have been
used to explain the heat capacity anomaly at 100 K, the hidden order
state, and the anomalous thermal expansion at 60 K need to be
re-examined within this framework. Our measurements show that the
first excited state, if allowed by selection rules, must exist above
200 K.

The role that these excitations play with the formation of the heavy
fermion state can be examined through their temperature dependence.
Figure 5 shows recent heat capacity measurements, in comparison with
inelastic neutron scattering scans at 100 K.  Note that the
incommensurate features have disappeared at this temperature, often
called the ``coherence temperature,'' which marks a cross-over from
paramagnetic weakly correlated moments to correlated heavy electron
behavior.  They are intimately correlated with the coherence
temperature and the formation of the heavy fermion state.  This
energy scale, as well, describes the temperature dependance of the
heat capacity anomaly at 17.5 K through fits to an activation
law,\cite{Palstra2} and the temperature dependence of the
incommensurate excitations below 17.5 K.\cite{Wiebe}  This is
further evidence that the gapping of these excitations is directly
related to the formation of the ``hidden order'' state.

In conclusion, our recent neutron scattering results in
URu$_2$Si$_2$
unambiguously demonstrate that itinerant-style excitations exist above {T$_0$%
~} at incommensurate wavevectors $\sim$ 0.4 a* from the
antiferromagnetic zone center. The gapping of these excitations
accounts for the change in $\gamma$, the electronic contribution of
the specific heat, through {T$_0$~}. The hidden order transition
appears to be a rearrangement of electrons at the Fermi surface in
an itinerant rather than localized electron picture.  Excitations
out of this state at these incommensurate wavevectors show a mode
softening which can perhaps be reminiscent of another dynamic ground
state which has no translational order but a prominent specific heat
anomaly - the superfluid liquid helium transition.\cite{Feynman}
Correlations between the heavy quasiparticles in URu$_2$Si$_2$ build
up below 100 K, and at 17.5 K, there is a transition to a new
condensate which still remains a mystery.  Even though cause of the
change in the electronic specific heat has been identified, the true
order parameter for this system has yet to be unveiled.

\begin{acknowledgments}
All correspondence can be directed to C.~R.~Wiebe at
cwiebe@magnet.fsu.edu.

The authors would like to acknowledge helpful discussions with
C.~Broholm, C.~D.~Batista, A.~Leggett, C.~M.~Varma, J.~S.~ Gardner,
J.~S.~Brooks, G.~S.~Boebinger, and B.~D.~Gaulin.  This work was made
possible by support through the NSF and the state of Florida.
G.~M.~Luke, G.~J.~MacDougall, and W.~J.~L.~Buyers acknowledge
support through NSERC and the CIAR.  The authors are grateful for
the local support staff at the NIST Center for Neutron Research.
Data analysis was completed with DAVE, which can be obtained at
http://www.ncnr.nist.gov/dave/.  The work at NIST is supported in
part by the National Science Foundation under Agreement No.
DMR-0454672.
\end{acknowledgments}

\section{Competing financial interests}
The authors declare that they have no competing financial interests.

\section{Experimental Methods}

Our measurements were made with a single crystal grown at McMaster
University grown by the Czochralski method using elemental U, Ru,
and Si, followed by annealing in argon at 900 degrees C. The
magnetic properties were confirmed through DC magnetometry
measurements on a Quantum Design SQUID, and heat capacity and
resistivity measurements on a Quantum Design PPMS. The neutron
scattering measurements were performed at the Disk Chopper
Spectrometer (DCS) at the NIST Center for Neutron Research in
Gaithersburg, Maryland. The 11.5 g URu$_2$Si$_2$ crystal was aligned
in the H0L plane and placed in a standard ILL cryostat and
measurements were taken using cold neutrons of 2.5 \AA~ and 5 \AA.
 The identification of the equipment is not intended to imply
recommendation or endorsement by the National Institute of Standards
and Technology, nor is it intended to imply that the equipment is
necessarily the best available for the purpose.

Scattering experiments were completed by C.~R.~W., J.~A.~J.,
G.~J.~M., H.~D.~Z., Y.~Q., J.~R.~D.~C., Z.~Y., and W.~J.~L.~B. The
crystals were grown by J.~D.~G. and G.~M.~L. Specific heat
measurements were made by Y.~J.~J. and L.~B.  Data analysis and
writing of the paper was completed by C.~R.~W., J.~A.~J., G.~J.~M.,
G.~M.~L., L.~B., and W.~J.~L.~B.

\clearpage

\begin{figure}[t]
\linespread{1}
\par
\begin{center}
\includegraphics[scale=0.8,angle=0]{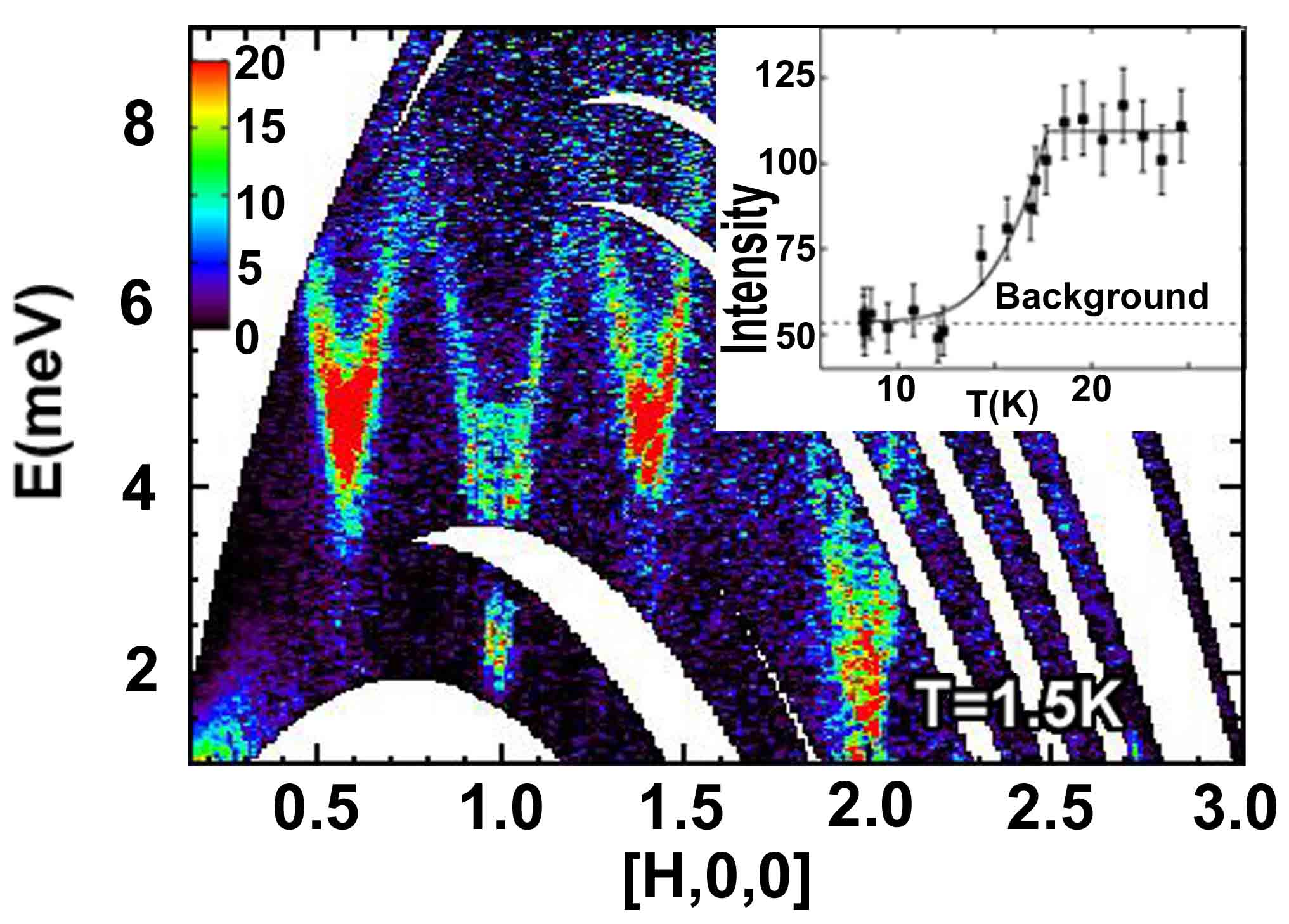}
\end{center}
\par
\linespread{1.6} \caption{Inelastic neutron scattering of
URu$_2$Si$_2$ in the H00 plane at T = 1.5 K (where H and L are
reciprocal lattice vectors a* and c*, L is integrated from -0.12 to
0.12). Note the minima at the AF zone center (100) and
incommensurate positions (1+/-0.4,0,0). The feature at (200) is due
to phonons. The inset shows how the incommensurate excitations
become gapped through the transition by counting at the point
(0.6,0,0) at 0.25 meV transfer on a triple axis
spectrometer./cite(Wiebe)} \label{fig1}
\end{figure}

\clearpage
\bigskip

\begin{figure}[tbp]
\linespread{1}
\par
\begin{center}
\includegraphics[scale=0.8,angle=0]{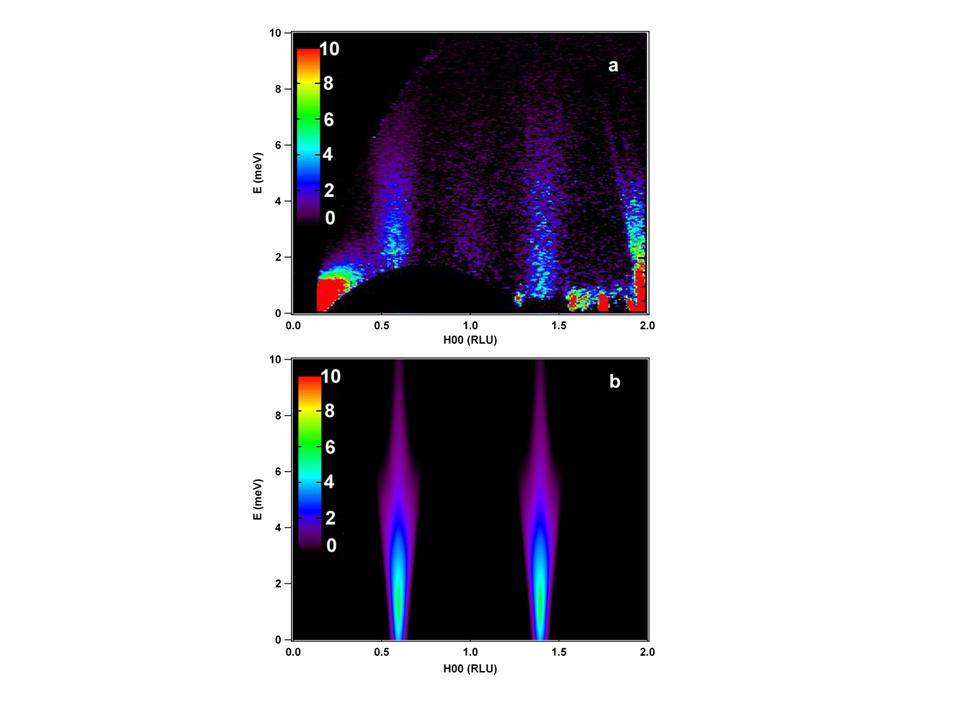}
\end{center}
\par
\linespread{1.6} \caption{Itinerant spin excitations in
URu$_2$Si$_2$ (a) Inelastic neutron scattering of URu$_2$Si$_2$ in
the H00 plane at T = 20 K (L integrated from -0.12 to 0.12). Note
the cone of ungapped excitations emanating from the incommensurate
wavevectors (1 +/- 0.4, 0, 0). (b)  Fits to the Chou model, with
parameters described in the text.} \label{fig2}
\end{figure}

\clearpage
\bigskip

\begin{figure}[t]
\linespread{1}
\par
\begin{center}
\includegraphics[scale=0.3,angle=0]{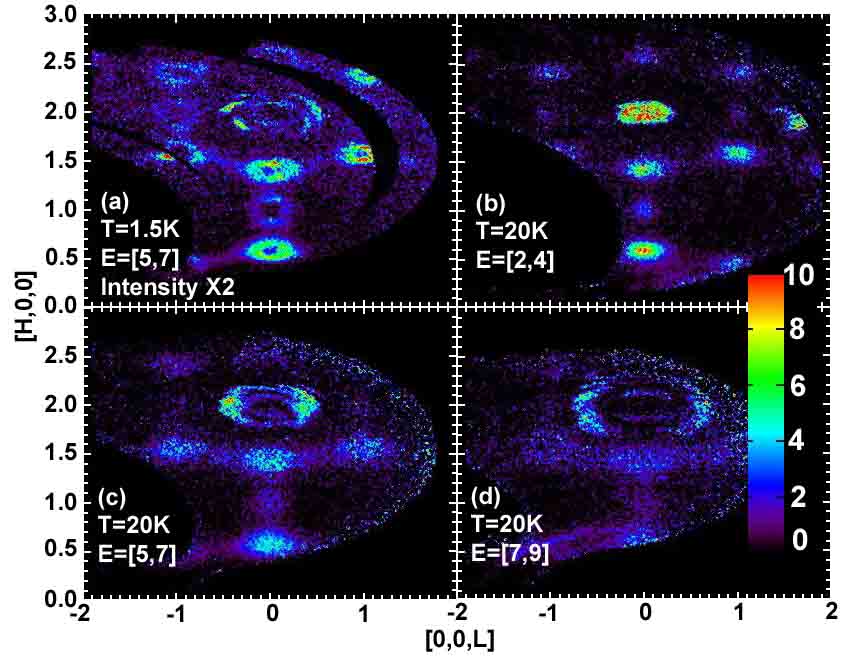}
\end{center}
\par
\linespread{1.6} \caption{Cuts along $\Delta$E of scattering in the
H0L plane at (a) 1.5 K and (b), (c), (d) 20 K. The rings about (200)
correspond to phonons. Note the spin waves in (a) as rings centered
on (H+K+L)= odd integers (AF zone center) and rings centered about
incommensurate points separated by 0.4 a* from the AF zone centers.
At 20 K, there are weak overdamped AF fluctuations at the zone
centers, but it is clear that a cone of highly correlated
excitations develops as a function of energy. The energy of
integration is denoted in meV for each plot.} \label{fig3}
\end{figure}

\clearpage
\bigskip

\begin{figure}[t]
\linespread{1}
\par
\begin{center}
\includegraphics[scale=0.5,angle=0]{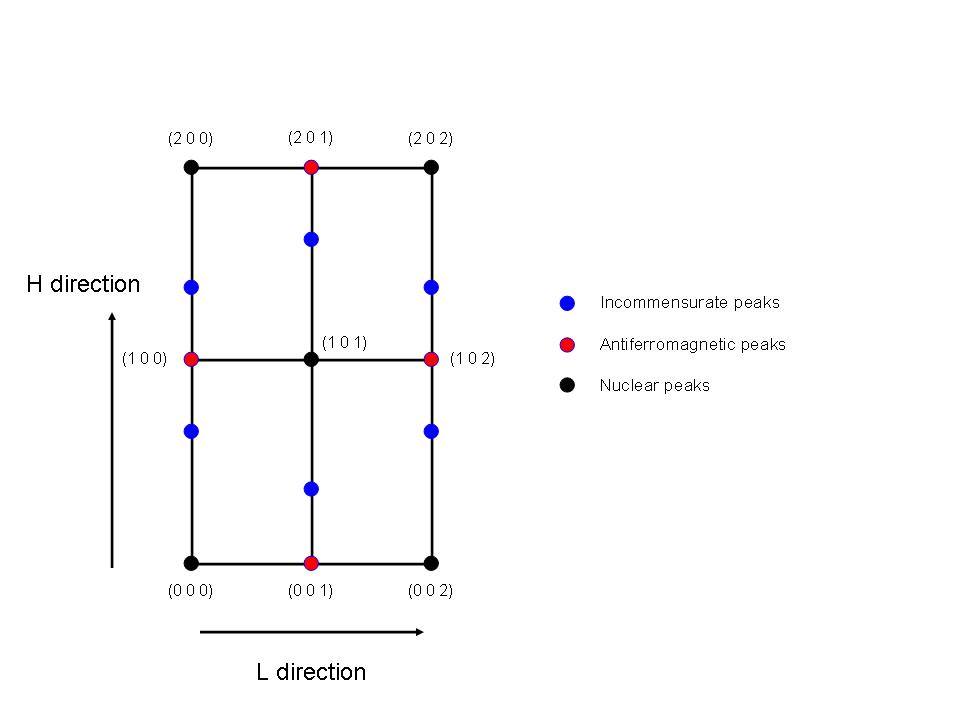}
\end{center}
\par
\linespread{1.6} \caption{Reciprocal space map of the commensurate
and incommensurate scattering in URu$_2$Si$_2$ (blue dots:
incommensurate positions, red dots: antiferromagnetic lattice
points, blue dots: nuclear lattice points} \label{fig5}
\end{figure}

\clearpage
\bigskip

\begin{figure}[t]
\linespread{1}
\par
\begin{center}
\includegraphics[scale=0.3,angle=0]{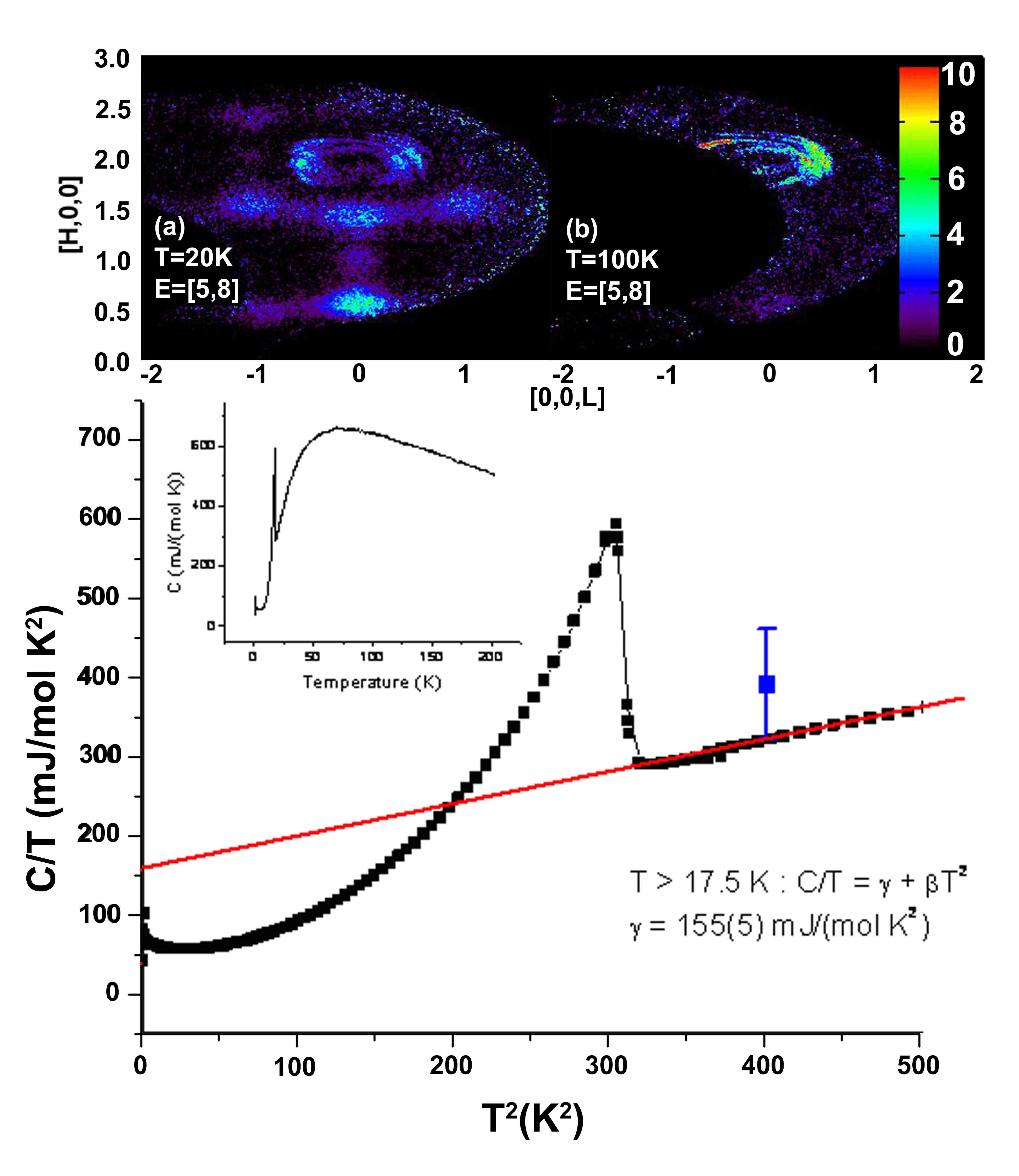}
\end{center}
\par
\linespread{1.6} \caption{Cuts in the H0L plane integrated from
5-8 meV at (a) 20 K and (b) 100 K. Note how the incommensurate
scattering at positions such as (1.6,0,1) disappears at 100 K
(indicated by a circle).  This is just above the ``coherence''
temperature where a signature of heavy quasiparticle formation is
seen in the specific heat (inset).(c) The specific
heat/temperature as a function of T$^2$, showing the anomaly at
17.5 K.  The linear portion of the specific heat, $\gamma$, is
calculated with the solid line to be 155(5) mJ/(mol K$^2$).  The
blue data point is the calculation of our $\gamma$ = 220(70)
mJ/(mol K$^2$) from the spin fluctuations observed at 20 K.  The
errors bars have been calculated from equation (2) in the text.}
\label{fig4}
\end{figure}

\end{document}